\documentclass[12pt,a4paper]{article}
\usepackage{amsfonts}
\title{A Novel Approach\\ to Non--Hermitian Random Matrix Models}
\author{Andrzej Jarosz  and  Maciej A. Nowak \\ \\ Marian Smoluchowski Institute of Physics,\\Jagiellonian University,\\Reymonta 4, 30--059 Krak\'ow, Poland
\\ \\\texttt{jarosz@th.if.uj.edu.pl} \\\texttt{nowak@th.if.uj.edu.pl}}
\begin{document}
\newcommand{\Det}{\mathrm{Det}}
\newcommand{\Tr}{\mathrm{Tr}}
\newcommand{\bTr}[1]{\mathrm{bTr}_{#1}}
\newcommand{\D}{\mathrm{D}}
\newcommand{\U}{\mathrm{U}}
\newcommand{\I}{\mathrm{I}}
\newcommand{\be}{\begin{eqnarray}}
\newcommand{\ee}{\end{eqnarray}}
\maketitle



\section*{Abstract}
\label{s:a}

In this paper we propose a new method for studying spectral properties
of the non-hermitian random matrix ensembles. 
Alike {\em complex} Green's function encodes, via discontinuities, 
the real spectrum 
of the hermitian ensembles, the proposed here {\em quaternion} 
extension of the 
Green's function  leads directly to complex spectrum in case of non-hermitian 
ensembles and encodes additionally
some spectral properties of the eigenvectors.
The standard two-by-two matrix representation 
of the quaternions leads to generalization of so-called  matrix-valued
resolvent, proposed recently in the context of diagrammatic 
methods~\cite{JNPZ1,JNPZ2,FZ1,FZ2,CW,FS}.
We argue that quaternion Green's function obeys Free Variables Calculus
~\cite{VOICULESCU,SPEICHER}. In particular, 
 the quaternion
functional inverse of the matrix Green's function, called
after~\cite{ZEE} Blue's function obeys simple addition law, as observed 
some time ago~\cite{JNPZ1,FZ1}. 
 Using this law  we
derive new,  general, algorithmic and efficient  method to
find the non--holomorphic Green's  function for {\em all} 
non-hermitian ensembles of the form $H + iH^{'}$, 
where ensembles $H$ and $H^{'}$ are  independent (free in the sense
of Voiculescu~\cite{VOICULESCU}) hermitian ensembles from arbitrary measure.
We demonstrate the power  of the method 
by a straightforward rederivation of spectral properties  
for several examples of
non--hermitian random matrix models.\\
\noindent
PACS: 05.40.+j; 05.45+b; 05.70.Fh;11.15.Pg\\
Keyword: Non--hermitian random matrix models.\\


\section{Introduction}
\label{s:b}
Random matrix models provide a powerful framework for modeling 
numerous physical phenomena, with  applications covering 
all branches of theoretical physics~\cite{GENERALHER,GUHR,ZINNJUSTIN}.
Among different classes of random matrix models the non-hermitian
random ensembles form a fascinating  class. Contrary to the hermitian 
ensembles, where real eigenvalues form cuts on real axis, 
general non-hermitian ensembles develop spectrum, which covers 
two-dimensional, often multiple-connected  support on the whole 
complex plane.
From mathematical point of view, non-hermitian ensembles are challenging, 
since several standard methods of hermitian random matrix calculus fail
in this case. On the other side, the issue of non-hermitian random matrix 
ensembles is far from being academic. Non-hermitian ensembles
are omnipotent in various branches of  physics and in interdisciplinary 
sciences. Sample applications are: 
open chaotic scattering~\cite{HAAKE}, spectral properties
of Euclidean Dirac operators in the presence
of chemical potential~\cite{CHEM} or CP-violating angle $\theta$ in
 Quantum Chromodynamics~\cite{THETA}, 
non-hermitian generalizations of Anderson 
localization in mesoscopic systems~\cite{HATANO}, 
modeling of the chemical transitions in dissipative systems~\cite{EWA},     
matrix generalizations of multiplicative diffusion processes~\cite{OURRECENT}
or evolution of the spectral curves for non-hermitian ensembles in the 
context of the growth problem~\cite{TEODOR}.
All these applications call for new calculational methods to deal 
with the problem of complex spectra. 
 
In this paper we present a  new technique, introducing the  quaternion 
generalization of the Green's function, the quaternion generalization 
of its functional inverse (quaternion Blue's function) and we formulate 
the quaternion addition law for non-hermitian ensembles.  
In section~2, we briefly remind  basic known facts on hermitian 
and non-hermitian Green's functions.
In section~3, we make a connection to Free Random Variables calculus. 
We comment also on some previous approaches based on diagrammatic
techniques. 
In section~4, after establishing the notation, we outline the construction 
and we present 
the quaternion generalization of the hermitian Green's  function
and its properties. In particular, we present a general form 
of the addition law for non-hermitian ensembles. 
   
In section~5, we formulate the operational form of the addition algorithm.
In section~6, we adapt  the operational form 
of the addition formalism for the particular case of the Gaussian 
randomness. Then, to demonstrate the power of the method, we 
provide  straightforward derivations of three classical results 
in non-hermitian random matrix models. 
Section~7 summarizes our conclusions. 
Appendices hide some necessary technical details and/or proofs. 
\section{Hermitian and non-hermitian Green's functions}
\def\no{\nonumber}
\def\qu{\quad}
\def\qb{\bar{q}}
\def\qbm{\bar{\mbox{q}}}
\def\la{\langle}
\newcommand{\gm}{\gamma}
\newcommand{\al}{\alpha}
\renewcommand{\th}{\theta}
\newcommand{\Sg}{\Sigma}
\newcommand{\dl}{\delta}
\newcommand{\SSg}{\tilde{\Sigma}}
\newcommand{\eq}{\begin{equation}}
\newcommand{\eqx}{\end{equation}}
\newcommand{\eqn}{\begin{eqnarray}}
\newcommand{\eqnx}{\end{eqnarray}}
\newcommand{\ben}{\begin{eqnaray}}
\newcommand{\een}{\end{eqnarray}}
\newcommand{\f}[2]{\frac{#1}{#2}}
\newcommand{\ra}{\longrightarrow}
\newcommand{\GG}{{\cal G}}
\renewcommand{\AA}{{\cal A}}
\newcommand{\GR}{G(z)}
\newcommand{\MM}{{\cal M}}
\newcommand{\BB}{{\cal B}}
\newcommand{\ZZ}{{\cal Z}}
\newcommand{\DD}{{\cal D}}
\newcommand{\HH}{{\cal H}}
\newcommand{\RR}{{\cal R}}
\newcommand{\GT}{{\cal G}_1 \otimes {\cal G}_2^T}
\newcommand{\GGb}{\bar{{\cal G}}^T}
\newcommand{\Du}{{\cal D}_1}
\newcommand{\Dl}{{\cal D}_2}
\newcommand{\zb}{\bar{z}}
\newcommand{\trqq}{\tr_{q\bar{q}}}
\newcommand{\arr}[4]{
\left(\begin{array}{cc}
#1&#2\\
#3&#4
\end{array}\right)
}
\newcommand{\arrd}[3]{
\left(\begin{array}{ccc}
#1&0&0\\
0&#2&0\\
0&0&#3
\end{array}\right)
}

\newcommand{\cor}[1]{\left\langle{#1}\right\rangle}
\newcommand{\ket}[1]{\left|{#1}\right\rangle}

\newcommand{\tr}{\mbox{\rm tr}\,}
\newcommand{\One}{\mbox{\bf 1}}
\newcommand{\pauli}{\sg_2}
\newcommand{\corr}[1]{\la{#1}\rangle}
\newcommand{\br}[1]{\overline{#1}}
\newcommand{\phib}{\br{\phi}}
\newcommand{\psib}{\br{\psi}}
\newcommand{\lm}{\lambda}
\newcommand{\ksi}{\xi}

\newcommand{\Gb}{\br{G}}
\newcommand{\Vb}{\br{V}}
\newcommand{\Gm}{G_{q\br{q}}}
\newcommand{\Vm}{V_{q\br{q}}}

\newcommand{\ggd}[2]{\GG_{#1}\otimes\GG^T_{#2}\Gamma}
\newcommand{\noi}{\noindent}

\newcommand{\qqqq}{\quad\quad\quad\quad}

Let us first focus on a hermitian
random matrix model. 
A basic tool to investigate eigenvalues' distribution for $H$ is a
resolvent, or Green's function,
\begin{equation} \label{eq:mmmmm}
G_{H}(z) = \frac{1}{N} \langle \Tr \frac{1}{z1_{N} - H} \rangle .
\end{equation}
where $1_N$ is a unit matrix of size $N$. 
The usefulness of this function stems from the fact, 
that, for any finite $N$,  $G_{H}(z)$ can
be written as
\begin{equation} \label{eq:nnnnn}
G_{H}(z) = \frac{1}{N} \langle \sum_{i = 1}^{N} \frac{1}{z -
\lambda_{i}} \rangle, \qquad \lambda_{i} \in \mathbb{R},
\end{equation}
after diagonalizing $H$ by a unitary similarity transformation, so
that $G_{H}(z)$ is a meromorphic function with $N$ poles at
$\lambda_{i}$ on the real line. In the large $N$ limit
 the poles merge into cuts on the real axis.

Green's function $G_{H}(z)$ can be used to reconstruct 
the spectral density function $\rho_{H}
(\lambda)$ due to the relation
\begin{equation} \label{eq:vvvvv}
\frac{1}{\lambda + i \epsilon} = {\rm PV} \frac{1}{\lambda} - i
\pi \delta (\lambda)
\end{equation}
where PV denotes the principal value distribution and 
$\epsilon$ is meant  implicitly  approaching
limit $\epsilon \to 0$, so that
\begin{equation} \label{eq:ooooo}
\rho_{H} (\lambda) = \frac{1}{N} \langle \Tr \delta (\lambda 1_{N}
- H) \rangle = -\frac{1}{\pi} \mathrm{Im} G_{H} (\lambda + i
\epsilon).
\end{equation}
Hence $\rho_{H} (\lambda)$ can be read out from discontinuities of
the imaginary part of the Green's function.

There are several ways of calculating Green's functions for 
hermitian random matrix models~\cite{GUHR,GENERALHER,ZINNJUSTIN}.
Here we mention  the diagrammatic approach, after~\cite{BZ}.
A starting point of the approach is the expression allowing for the
reconstruction of the Green's function from all the moments.
\begin{equation} \label{eq:ppppp}
G_{H}(z) = \sum_{n \geq 0} \frac{m_{H, n}}{z^{n + 1}}, \qquad
m_{H, n} = \frac{1}{N} \langle \Tr H^{n} \rangle .
\end{equation}
The resolvent $G_{H}(z)$ is interpreted as a series in the guise 
of  the Feynman-like  diagrammatic expansion in the
large number of colors limit.
This series can be ef\mbox{}ficiently evaluated exploiting the
analogy of RMT to $0 + 0$ dimensional gauge field theory in the 't
Hooft large $N$ number of colors limit~\cite{THOOFT}. To avoid 
unnecessary repetitions, we refer to the literature~\cite{JNPZ2,BZ}.

The reason why the above procedure works correctly for {\em hermitian}
matrix models is the fact that the Green's function is guaranteed to
be {\em holomorphic} in the whole complex plane except at most on one or
more 1-dimensional intervals.

The key dif\mbox{}ference which arises in the non--hermitian case
(we denote the general non-hermitian matrix by $X$) 
is that eigenvalues of $X$ are complex in general; in the large
$N$ limit they form two--dimensional domains in the complex plane,
in contrary to one--dimensional cuts in the previous case.
 Therefore the power series expansion (\ref{eq:ppppp}) no
longer captures the full information about the Green's function.
In particular the eigenvalue distribution is related to the
non-analytic (non-holomorphic) behavior of the Green's function:
\eq
\label{e.dzbar}
\f{1}{\pi}\partial_{\zb} G(z)=\rho(z) \ .
\eqx

This phenomenon can be easily seen even in the the simplest
non-hermitian ensemble --- the Ginibre-Girko one \cite{GinGir,GIRKO},
with non-hermitian matrices $X$, and measure
\be
P(X)= e^{-N {\rm Tr} X X^{\dagger}} \,.
\ee
It is easy to verify that all moments vanish $\cor{\tr X^n}=0$, for $n>0$
so the expansion (\ref{eq:ppppp}) gives the Green's function to be
$G(z)=1/z$. The true answer is, however, different. Only for $|z|>1$
one has indeed $G(z)=1/z$. For $|z|<1$ the Green's function is
nonholomorphic and equals $G(z)=\zb$.

The above difficulty was first addressed in mathematical papers.
Brown~\cite{Bro86} defined a measure for complex ensembles as 
\be
\mu_X=\frac{1}{2\pi} \left( \frac{\partial^2}{(\partial \Re \lambda)^2}
+ \frac{\partial^2}{(\partial \Im \lambda)^2}\right) \log
\det(X-\lambda 1_N)
\label{Brownmeasure}
\ee
where 
\be
\det (X-\lambda 1_N)= \exp \left[\frac{1}{N} {\rm Tr} \log 
\sqrt{(X-\lambda 1_N)(X^{\dagger}-\bar{\lambda} 1_N)}\right]
\ee
is known in mathematics as Fuglede-Kadison determinant. 
For some recent results on Brown measure we refer to \cite{HL00,Sni02}.  

Physicists have addressed the problem of measure, 
exploiting the analogy to two-dimensional electrostatics
\cite{HAAKE,GIRKO,SOMMERS}. 
Let us define the ``electrostatic potential''
\be
F&=&\frac{1}{N} {\rm Tr } \ln [( z 1_N-X)(\bar{z}1_N-X^{\dagger}) +
        \epsilon^2 1_N] \,.\nonumber \\
&=& \frac{1}{N} \ln {\rm det} [( z1_N -X)(\zb 1_N-X^{\dagger}) +
        \epsilon^2 1_N]
\label{els}
\ee
Then
\eqn
\lim_{\epsilon \rightarrow 0} \frac{\partial^2 F(z,\bar{z})}%
        {\partial z\partial \bar{z}}
&=&\lim_{\epsilon \rightarrow 0}
\frac{1}{N}\left\langle\!{\rm Tr} \frac{\epsilon^2}{(|{ z1_N}\!-\!X|^2
+\!\epsilon^2 1_N)^2}\!\right\rangle
\nonumber \\
 &=&
    \frac{\pi}{N} \left\langle \sum_i \delta^{(2)}(z\!-\!\lambda_i)
        \right\rangle
\equiv
\pi \rho(x,y)
\label{spec}
\eqnx
represents Gauss law,
where $z=x+iy$. The last equality follows from 
the representation of the complex Dirac delta
\be
\pi \delta^{(2)}(z-\lm_i)=\lim_{\epsilon \rightarrow 0} 
\frac{\epsilon^2}{(\epsilon^2 +|z-\lm_i|^2)^2}
\ee
In the spirit of the electrostatic analogy we can define
the Green's function $G(z,\bar{z})$, as an ``electric field''
\be
G
\equiv\!
 \frac{\partial F}{\partial z}\!=\!
\frac{1}{N}\!\lim_{\epsilon \rightarrow 0}\!
\left\langle\!{\rm Tr}\!
\frac{\bar{z} 1_N-X^{\dagger}}{(\bar{z} 1_N\!-\!X^{\dagger})({ z}1_N\!-\!X)
+\!\epsilon^2 1_N)}\!\right\rangle .\!
\label{GG}
\ee
Then Gauss law leads (\ref{e.dzbar}). 

The drawback of the above construction is that 
due to the quadratic structures in the denominator, 
form (\ref{GG}) is  difficult to use in practical calculation. 

Instead of working {\it ab initio} with the object
(\ref{GG}), one can take consider the following generalization
~\cite{JNPZ2}.
One  defines the matrix-valued resolvent
through
\be
\hat{{\cal {G}}}
&=&\frac{1}{N} \left\langle \rm{bTr_{2}}
\setlength\arraycolsep{0pt}
\arr{{z}1_N-X}{i  \epsilon 1_N}{i  \epsilon 1_N}{{\zb}1_N
-X^{\dagger}}^{-1}\right\rangle = \nonumber\\
&=&
\f{1}{N} \left\langle {\rm bTr_{2}} 
\arr{A}{B}{C}{D} \right\rangle \equiv
\setlength\arraycolsep{3pt}
\arr{{\GG}_{11}}{{\GG}_{1\overline{1}}}{{\GG}_{\overline{1}1}}
{{\GG}_{\overline{1}\overline{1}}}
\label{19}
\ee
with 
\be 
A&=&\frac{{ \zb 1_N} -X^{\dagger}}{({
          \zb 1_N} -X^\dagger)({ z}1_N-X)+\epsilon^2 1_N} \nonumber \\
B&=&\frac{-i\epsilon }{({
          z}1_N -X)({ \zb}1_N-X^{\dagger})+\epsilon^2 1_N} \nonumber \\
C&=&\frac{ -i\epsilon}{({
          \zb}1_N -X^\dagger)({z}1_N-X)+\epsilon^2 1_N} \nonumber \\
D&=&\frac{ {z}1_N -X}{({
          z}1_N -X)({ \zb}1_N-X^{\dagger})+\epsilon^2 1_N} 
\ee
and 
where we introduced the `block trace'  defined as
\be
{\rm bTr_{2}}
\setlength\arraycolsep{3pt}
\arr{A}{B}{C}{D}_{2N \times 2N} \hspace*{-3mm}\equiv
\setlength\arraycolsep{3pt}
\arr{
{\rm Tr}\ A}{{\rm Tr}\ B}{{\rm Tr}\ C}{{\rm Tr}\ D}_{2 \times 2}
\hspace*{-5mm}\,.
\label{block2}
\ee
Then, by definition, the upper-right component $\GG_{11}$, hereafter
denoted by $G_X(z,\zb)$, is equal to
the Green's function (\ref{GG}).

The block approach has several advantages. First of all it is {\em
linear} in the random matrices $X$ allowing for a simple
diagrammatic calculational procedure.
Let us define $2N$ by $2N$ matrices
\be
\label{defzg}
Z_{\epsilon}=
\arr{{z 1_N}}{i \epsilon { 1_N}}{i 
 \epsilon { 1_N}}{\bar{z} 1_N} \quad, \quad
\HH=\arr{X}{0}{0}{X^{\dagger}} \,.
\ee
Then the generalized Green's function is given formally by the same
definition as the usual Green's function $G$,
\be
\GG=\frac{1}{N} \left\langle {\rm bTr_{2}} 
\frac{1}{Z_{\epsilon}-\HH}\right\rangle
        \,.
\label{concise}
\ee
What is more important, also in this case the Green's function is completely
determined by the knowledge of all  matrix-valued moments
\be
\left\langle {\rm bTr_{2}}\,\,\, {Z_{\epsilon}}^{-1} \HH 
{Z_{\epsilon}}^{-1}
 \HH \ldots {Z_{\epsilon}}^{-1}
        \right\rangle \,.
\label{genmom}
\ee
This last observation allowed for a diagrammatic interpretation~\cite{JNPZ2}. 
The
Feynman rules were analogous to the hermitian ones, only now one has to
keep track of the block structure of the matrices.

Let us summarize the  general properties~\cite{JNPZ2} of the matrix valued
generalized Green's function. Each component of the matrix carries important
information about the stochastic properties of the system.
There are always two solutions for $\GG_{11}(X)$, one holomorphic, another
non-holomorphic. The second one leads, via Gauss law, to the eigenvalue
distribution. The first one, the holomorphic one is not ``spurious'', 
it represents the generating function for the real {\em moments} of the 
complex distribution~\cite{NEWPAPER}, but, by definition, cannot reproduce
the non-holomorphic spectrum of the ensemble.  

The shape of the ``coastline'' bordering the ``sea'' of
complex eigenvalues is determined by the
matching conditions for the two solutions,
i.e. it is determined by imposing on the non-holomorphic solution 
the condition $\GG_{12}\GG_{21}=0$.
The off-diagonal elements have an interesting interpretation~\cite{USEVECT}.
They represent the correlator between the left and right 
{\em eigenvectors} (introduced in ~\cite{MEHLIG})
\be
\left< \sum_a (L_a|L_a)(R_a|R_a)\delta(z-\lambda_a)\right>
=-\frac{N}{\pi}\GG_{12}\GG_{21}\equiv -\frac{N}{\pi} C_X(z,\bar{z}) 
\label{eigenveccorr}
\ee
On boundary of the domain of the eigenvalues, the above correlator vanishes. 

We would like to mention for completeness, that another, similar approach
appears often in the literature, under the ``hermitization 
method'' name~\cite{FZ1,FZ2,CW}.
Basically it uses an  alternative representation for the  determinant 
in (\ref{els}), which  is rewritten in a  hermitian form
\be
{\rm det}[(z 1_N -X)(\bar{z} 1_N-X^{\dagger}) +\epsilon^2 1_N]=
- {\rm det}
\setlength\arraycolsep{0pt}
\arr{1_N}{z 1_N -X}{\zb 1_N
-X^{\dagger}}{-\epsilon^2 1_N}
\ee
Both versions lead to the similar results.

An important feature of the explained above 
method of the generalized matrix-valued Green's functions
 is  their link to Free Random Variables
calculus~\cite{VOICULESCU,SPEICHER}, which we will 
address in the next section.

\section{Free Random Variables Approach}

In the hermitian RMT a method of free random variables (FRV),
introduced by Voiculescu~\cite{VOICULESCU}, is widely 
used~\cite{ZEE,SPEICHERNEU,
GROSS,BLUESHADES}.
The power of the method lies in the fact, that in particular it allows
to find the resolvent for the sum of two 
independent
 (free)~\footnote{Freeness is a non-commutative analog of 
statistical independence,
introduced by Voiculescu~{\protect\cite{VOICULESCU}}.}
ensembles $H_1$ and $H_2$, 
on the basis of knowledge of the resolvents for each
separated ensemble only. In other words, operation R introduced
by Voiculescu  linearizes the non-commutative, matrix convolution 
of hermitian random variables, alike the logarithm of characteristic function
does it for sum  of  two identical, independent, one-dimensional 
random variables. 

In this paper  we recall a version of Free Random Variables 
formulated  by Zee~\cite{ZEE}  and known under the name  
\emph{Blue's functions}.
Blue's function is simply related to original R-function 
of Voiculescu
\be
B(z)=R(z)+\frac{1}{z}
\ee
and its usefulness comes from the fact, 
that it is simply the functional inverse of the resolvent
\begin{equation} \label{eq:xxxxx}
B(G(z)) = G(B(z)) = z.
\end{equation}
The algorithm of ``adding'' ensembles is therefore
straightforward:  
 consider two freely independent hermitian random
matrices, $H$ and $H'$, and their (holomorphic) Green's functions,
$G_{H}(z)$ and $G_{H'}(z)$.  The addition
algorithm goes as follows: First, one inverts 
functionally $G_{H}(z)$ and $G_{H'}(z)$
 to get Blue's  functions $B_{H}(z)$ and $B_{H'}(z)$.
Second, using the addition law 
\begin{equation} \label{eq:yyyyy}
B_{H + H'}(z) = B_{H}(z) + B_{H'}(z) - \frac{1}{z}.
\end{equation}
one obtains the Blue's function for the sum of ensembles.
Third, inverting  functionally this function 
one recovers needed  $G_{H + H'}(z)$.

The generalization  of the concept of the Blue's function
for non-hermitian ensembles was first proposed in ~\cite{JNPZ1} and 
confirmed by ~\cite{FZ1,FZ2}.
The generalized, matrix-valued Blue's function was defined
as  a functional inverse of matrix-valued Green's function, 
i.e. 
\be 
\GG(\BB(Z_{\epsilon}))=Z_{\epsilon}
\label{mblue}
\ee
for matrix $Z$ (\ref{defzg}).
Such defined Blue's functions fulfilled
the similar, but now matrix-valued addition law, i.e. 
\be
\BB_{1+2}(Z)=\BB_1(Z)+\BB_2(Z)-Z^{-1} 
\ee
with $\epsilon$ safely put to zero. 
The explicit form of the matrix-valued Blue's function for an arbitrary 
ensembles 
was, however, far from obvious. In several applications using this method, 
an additional insight coming from e.g. diagrammatic interpretation was 
necessary.  

One of the  purposes of this paper is 
to propose the rigorous method yielding the matrix-valued
Blue's function solely on the basis of algebraic properties, 
and abandoning any additional insight from e.g. 
diagrammatization methods. 
We will shown in the following chapters, that such construction is 
not only possible, but leads to strikingly simple 
algorithms for addition of a broad  class of non-hermitian ensembles. 

\section{Quaternion Green's functions}

The main advantage of working with the complex Green's function
for hermitian ensembles stems from the  holomorphic properties
 of the resolvent. 
This function, being holomorphic  everywhere except 
the intervals on real axis, allows, by analytical calculation, 
to recover  the spectral function by approaching the cuts from 
above and from below, i.e. by calculating the  discontinuities along the cuts.

It is tempting to propose the similar method in the case
of complex spectra. A natural generalization is the algebra of 
quaternions. Such a speculation appeared in 
the literature~\cite{FZ2}, 
but the explicit realization of such construction was never completed 
and left as a challenge.
In this chapter we propose the solution.   
We will now proceed in the following way:
\begin{enumerate}
\item First, we introduce the notation.
\item Second, we generalize the notion of a matrix--valued Green's
function (\ref{concise}) by exchanging $Z_{\epsilon}$ by a
general quaternion $Q$ thus exploiting the whole four--dimensional
quaternion space and not only its part 'close' to the complex
plane. A quaternion Blue's function is its functional inverse.

\item Third, we compute quaternion Green's and Blue's functions for a
hermitian random matrix $H$.

\item Fourth, we investigate how these functions behave during
multiplying a general random matrix by a fixed complex number $g$.

\item Fifth, we present an addition formula which enables to
express a quaternion Blue's function of a sum of two freely
independent general random matrices by their respective quaternion
Blue's functions. Then we compute a quaternion Blue's function for 
a non-hermitian $X = H +
iH'$, $H$ and $H'$ being two hermitian and freely independent
random matrices.

\end{enumerate}

\subsection{Notation}
\label{s:d}

We found it convenient to introduce the following notation:\\
Writing down explicitly any square $N \times N$ matrix we
add a subscript '$_{N}$' to make the reader sure about the
matrix' size. The blank places in a matrix mean that they
are occupied by zeroes. 
 We use standard  Pauli matrices. 
If $X$ is a general complex $N \times N$ matrix then
\begin{equation} \label{eq:e}
X^{\D} = \left( \begin{array}{cc} X & \\ & X^{\dagger} \end{array}
\right)_{2N},
\end{equation}
where the superscript '$^{\D}$' means '\underline{d}uplication'
since we duplicate $X$ into $X$ and $X^{\dagger}$.
For quaternions, i. e. combinations of $1_{2}$ and three Pauli
matrices of the form
\begin{equation} \label{eq:aaaaaa}
Q = x_{0}1_{2} + i \vec{x} \cdot \vec{\sigma}, \qquad x_{0},
x_{1}, x_{2}, x_{3} \in \mathbb{R}
\end{equation}
we use a $2 \times 2$ matrix notation
\begin{equation} \label{eq:a}
Q = \left( \begin{array}{cc} a & i\bar{b} \\ ib & \bar{a}
\end{array} \right)_{2}, \qquad a, b \in \mathbb{C}
\end{equation}
with $a = x_{0} + ix_{3}, b = x_{1} + ix_{2}$.
If $Q$ is a general quaternion, then
\begin{equation} \label{eq:b}
Q^{\U} = \left( \begin{array}{cc} a1_{N} & i\bar{b}1_{N} \\
ib1_{N} & \bar{a}1_{N} \end{array} \right)_{2N},
\end{equation}
where the superscript '$^{\U}$' means '\underline{u}niatal
extension', since we extend $Q$ by multiplying each of its
elements by a unit matrix ($N$ is a fixed number defined below)
and also
\begin{equation} \label{eq:c}
Q^{\I} = Q \left( \begin{array}{cc} i & \\ & -i \end{array}
\right)_{2} = Qi\sigma_{3} = \left( \begin{array}{cc} ia &
\bar{b} \\ -b & -i\bar{a} \end{array} \right)_{2},
\end{equation}
where the superscript '$^{\I}$' comes from \underline{i}maginary
units appearing in (\ref{eq:c}).
A quaternion of a special interest for us is defined when a
complex variable $z$ is being considered,
\begin{equation} \label{eq:d}
Z = \left( \begin{array}{cc} z & \\ & \bar{z} \end{array}
\right)_{2}.
\end{equation}
Generally, quaternion Green's (Blue's) functions we denote by 
calligraph alphabet to avoid confusion with  complex 
standard Green's (Blue's)
functions $G(z)$ ($B(z)$).
\subsection{Definition of a Quaternion Green's Function}
\label{s:ss}

Let us consider a random $N \times N$ matrix $X$, hermitian or
not. We see that the key quantity for $X$ is its matrix--valued
Green's function (\ref{concise}); it has the form of a complex
Green's function  with a complex variable $z$
substituted by a $2N \times 2N$ matrix $Z_{\epsilon}^{\U}$ and $X$
substituted by a $2N \times 2N$ matrix $X^{\D}$.

Let us consider a straightforward generalization of
(\ref{concise}): from now on we  call \emph{the quaternion
Green's function} the object
\begin{equation} \label{eq:f}
\mathcal{G}_{X}(Q) = \frac{1}{N} \langle \bTr{2} \frac{1}{Q^{\U} -
X^{\D}} \rangle,
\end{equation}
where $Q$ is a quaternion
\begin{equation} \label{eq:g}
Q = \left( \begin{array}{cc} a & i\bar{b} \\ ib & \bar{a}
\end{array} \right)_{2}.
\end{equation}
This defines $\mathcal{G}_{X}$ as a quaternion function of a
quaternion variable.

The only dif\mbox{}ference to the original meaning of this notion
is that we exchanged $Z_{\epsilon}$ with a general quaternion $Q$ thus
promoting $\mathcal{G}_{X}$ to be \emph{the function of $Q$}.
Particularly, for $Q = Z_{\epsilon}$ we arrive at the original
 meaning~\cite{JNPZ1}. The
key idea of this work  is that $\mathcal{G}_{X}$ has a few
new interesting features \emph{as a function of $Q$} which
cannot be seen when regarding only the previous case of $Q = Z_{\epsilon}$,
i.e. assuming that $\epsilon$ is infinitesimally small.

The first interesting feature of $\mathcal{G}_{X}$ is that it
enables to define \emph{the matrix Blue's function} as the
\emph{functional} inverse of $\mathcal{G}_{X}$:
\begin{equation} \label{eq:h}
\mathcal{G}_{X}(\mathcal{B}_{X}(Q)) =
\mathcal{B}_{X}(\mathcal{G}_{X}(Q)) = Q,
\end{equation}
for any $Q$. This is clearly the quaternion function of the quaternion
variable.
It is a natural generalization from the complex case
(\ref{eq:xxxxx}).
Note that 
if $Q$ is a quaternion, then $\mathcal{G}_{X}(Q)$ and
$\mathcal{B}_{X}(Q)$ are quaternions too.
 It requires $\mathcal{G}_{X}$ as a
function of $Q$ everywhere and not only 'near' $Z$.
It is particularly visible when studying the 
\emph{eigenvalues} of
$Q$. A  quaternion $Q$
(\ref{eq:aaaaaa}) has a very simple structure of eigenvalues,
\begin{equation} \label{eq:zzzzz}
q_{1} \equiv q = x_{0} + i|\vec{x}|, \qquad q_{2} = \bar{q} =
x_{0} - i|\vec{x}|.
\end{equation}
The eigenvalues have following basic  properties:
\begin{itemize}
\item They are mutually conjugated.
 \item They are dif\mbox{}ferent
(there is no degeneracy) always except of a very special case of
$\vec{x} = 0$, which means that the only degenerate quaternion is
the one of the form $Q = x_{0}1_{2}$. In other words, they
are non--real always except the degenerate case.
\item $|q|^{2} = x^{2}_{0} + \vec{x}^{2} = |a|^{2} + |b|^{2} = \Det Q$.
Since the determinant is positive, it always fulfills the 
requirement of Fuglede-Kadison construction for Brown measure.
\end{itemize}

Let us finish with  speculative remarks. Introducing quaternions
seems to be very natural from the algebraic point of view, however
the reader will find by inspection, that 
 for applications presented here not the
whole quaternion space is necessary; it can be restricted to the
subspace with
$x_{2} = 0$. 
Since  our calculations do not simplify considerably with  regards to
 $x_{2}
= 0$ case,  we assume the general case $x_{2} \neq 0$.
Finally, we mention that the construction 
presented here holds if  instead of quaternions 
a general $2 \times 2$ matrix (biquaternion) is  used.
In this case the matrix-space is eight-dimensional, however some 
nice properties of quaternion eigenvalues are lost.


\subsection{Quaternion Green's and Blue's  functions for a Hermitian Random
Matrix} \label{s:h}

Consider a known hermitian random $N \times N$
matrix $H$. We show how to calculate quaternion Green's
 and Blue's  functions for hermitian $H$.
The method is based upon a simple behavior of
quaternion Green's functions under similarity transformations
(\ref{eq:k}) and a knowledge of their form on diagonal matrices
(\ref{eq:i}). The same holds for Blue's functions, (\ref{eq:p})
and (\ref{eq:j}).

Let us first note, that a quaternion Green's function for $H$ at the
point of a \emph{diagonal matrix} has a very simple form of a
diagonal matrix with holomorphic Green's functions for $H$ on the
diagonal
\begin{equation} \label{eq:i}
\mathcal{G}_{H}(\left( \begin{array}{cc} q & \\ & \bar{q}
\end{array} \right)_{2}) = \left( \begin{array}{cc} G_{H}(q) &
\\ & G_{H}(\bar{q}) \end{array} \right)_{2}.
\end{equation}
Applying $\mathcal{B}_{H}$ to both sides of (\ref{eq:i}) and
exchanging $q$, $\bar{q}$ with $G_{H}(q)$, $G_{H}(\bar{q})$ we get
an identical theorem for a quaternion Blue's function for $H$:
\begin{equation} \label{eq:j}
\mathcal{B}_{H}(\left( \begin{array}{cc} q & \\ & \bar{q}
\end{array} \right)_{2}) = \left( \begin{array}{cc} B_{H}(q) &
\\ & B_{H}(\bar{q}) \end{array} \right)_{2}.
\end{equation}
To regain a general quaternion $Q$ from its eigenvalues we use a
similarity transformation $S$. The quaternion Green's function for
$H$ cooperates very well with similarity transformations of its
argument; consider a similarity transformation $S$ that is an
internal operation in the quaternion space, then
\begin{equation} \label{eq:k}
\mathcal{G}_{H}(S^{-1}QS) = S^{-1} \mathcal{G}_{H}(Q) S.
\end{equation}
The similar  formula holds for a quaternion Blue's function for
$H$,
\begin{equation} \label{eq:p}
\mathcal{B}_{H}(S^{-1}QS) = S^{-1} \mathcal{B}_{H}(Q) S.
\end{equation}
Proofs are presented in appendix \ref{s:kk}.

From these two properties we can also deduce  that eigenvalues
of $\mathcal{G}_{H}(Q)$ (resp. $\mathcal{B}_{H}(Q)$) are
$(G_{H}(q), G_{H}(\bar{q}))$ (resp. $(B_{H}(q),
B_{H}(\bar{q}))$).

These two simple algebraic properties of quaternion Green's and Blue's
functions, (\ref{eq:i}), (\ref{eq:k}) and (\ref{eq:j}),
(\ref{eq:p}) allow us to calculate them for any $Q$.
Consider now any non--degenerate quaternion $Q = \left( \begin{array}{cc} a & i\bar{b} \\
ib & \bar{a} \end{array} \right)_{2}$ with eigenvalues $q$,
$\bar{q}$ and let $S$ be the similarity transformation
diagonalizing $Q$,
\begin{equation} \label{eq:q}
Q = S^{-1} \left( \begin{array}{cc} q & \\ & \bar{q} \end{array}
\right)_{2} S.
\end{equation}
It is straightforward  
 to apply expressions (\ref{eq:i}) and
(\ref{eq:k}) to find a general form of $\mathcal{G}_{X}(Q)$. For a
given $Q$ we compute an explicit form of $S$ and $S^{-1}$, then a
simple three--matrix multiplication leads to the result. The result 
is derived 
in appendix \ref{s:ll}:
\begin{equation} \label{eq:w}
\mathcal{G}_{H}(Q) = \gamma_{H}(q, \bar{q}) 1_{2} - \gamma'_{H}(q,
\bar{q}) Q^{\dagger},
\end{equation}
where $\gamma_{H}$ and $\gamma'_{H}$ are two scalar functions
depending only on $Q$'s eigenvalues and are  given by
\begin{eqnarray}
\gamma_{H}(q, \bar{q}) & = & \frac{qG_{H}(q) - \bar{q}G_{H}(\bar{q})}{q - \bar{q}},  \label{eq:x}\\
\gamma'_{H}(q, \bar{q}) & = & \frac{G_{H}(q) - G_{H}(\bar{q})}{q -
\bar{q}}. \label{eq:y}
\end{eqnarray}
The degenerate case of $Q = x_{0}1_{2}$ is trivial  since the 
degenerate quaternion is already diagonal, so that from
(\ref{eq:i})
\begin{equation} \label{eq:bbbbbb}
\mathcal{G}_{H}(x_{0}1_{2}) = G_{H}(x_{0})1_{2}.
\end{equation}
This is the solution; we have expressed $\mathcal{G}_{H}$ by
$G_{H}$, i.e. the quaternion generalization using the 
known, holomorphic resolvent.


Let us note, that due to (\ref{eq:j}) and (\ref{eq:p}) the
identical construction can be made for $\mathcal{B}_{H}$; hence
the analogous expression for non--degenerate quaternion $Q$,
\begin{equation} \label{eq:z}
\mathcal{B}_{H}(Q) = \beta_{H}(q, \bar{q}) 1_{2} - \beta'_{H}(q,
\bar{q}) Q^{\dagger},
\end{equation}
where $\beta_{H}$ and $\beta'_{H}$ are similar to $\gamma_{H}$ and
$\gamma'_{H}$ but with $B_{H}$ in the place of $G_{H}$,
\begin{eqnarray}
\beta_{H}(q, \bar{q}) & = & \frac{qB_{H}(q) - \bar{q}B_{H}(\bar{q})}{q - \bar{q}},  \label{eq:aa}\\
\beta'_{H}(q, \bar{q}) & = & \frac{B_{H}(q) - B_{H}(\bar{q})}{q -
\bar{q}}. \label{eq:bb}
\end{eqnarray}
If a degeneracy is present, a formula analogous to
(\ref{eq:bbbbbb}) holds.
Again, we have expressed quaternion $\mathcal{B}_{H}$ by
standard  analytic $B_{H}$.


\subsection{Quaternion Green's and Blue's functions for a General
Random Matrix Multiplied by a Fixed Complex Number} \label{s:o}

We have already computed quaternion Green's and Blue's functions for a
hermitian random matrix $H$. Now there is time for the passage to
a non--hermitian random matrix. The following problem is the
introductory one before doing so: consider a general random matrix
$X$ and a fixed complex number $g$. We want to express
 the quaternion Green's
and Blue's  functions for $gX$ through the ones for $X$.

If $g = 0$ then trivially $\mathcal{G}_{0}(Q) = \frac{1}{Q}$, so
let us further assume that $g \neq 0$. In appendix \ref{s:mm}
we present a  simple derivation of
\begin{equation} \label{eq:ff}
\mathcal{G}_{gX}(Q) = \mathcal{G}_{X}( \left( \begin{array}{cc}
1/g & \\ & 1/\bar{g} \end{array} \right)_{2} Q ) \left(
\begin{array}{cc} 1/g & \\ & 1/\bar{g}
\end{array} \right)_{2}.
\end{equation}

This would be incorrect unless the argument were just  a quaternion $Q$; in
fact $\left( \begin{array}{cc} 1/g & \\ & 1/\bar{g}
\end{array} \right)_{2} Q$ is a quaternion with
\begin{equation} \label{eq:cccccc}
a \longrightarrow a/g, \qquad b \longrightarrow b/\bar{g}
\end{equation}
(see (\ref{eq:a})). In terms of variables  $x_i$, (see (\ref{eq:aaaaaa})), this
replacement  is non--trivial.

For a special case of real $g$, Eq.~(\ref{eq:ff}) simplifies,
\begin{equation} \label{eq:gg}
\mathcal{G}_{gX}(Q) = \frac{1}{g} \mathcal{G}_{X}( \frac{1}{g} Q),
\qquad g \in \mathbb{R}\setminus \lbrace 0 \rbrace.
\end{equation}
Let us note that this formula is identical to the one well known  for a
complex Green's function and complex $g$,
\begin{equation} \label{eq:hh}
G_{gX}(z) = \frac{1}{g} G_{X}( \frac{1}{g} z ).
\end{equation}
Similar behavior appears at the level of Blue's functions.  
Again, if  $g = 0$,  then trivially $\mathcal{B}_{0}(Q) = \frac{1}{Q}$, so
let us further assume that $g \neq 0$. From the definition
(\ref{eq:h}) and (\ref{eq:ff}) we immediately get
\begin{equation} \label{eq:ii}
\mathcal{B}_{gX}(Q) = \left( \begin{array}{cc} g & \\ & \bar{g}
\end{array} \right)_{2} \mathcal{B}_{X}( Q \left(
\begin{array}{cc} g & \\ & \bar{g} \end{array} \right)_{2} ).
\end{equation}
The argument here is a quaternion with
\begin{equation} \label{eq:dddddd}
a \longrightarrow ag, \qquad b \longrightarrow b \bar{g}.
\end{equation}
For a special case of real $g$ Eq.~(\ref{eq:ii}) simplifies,
\begin{equation} \label{eq:jj}
\mathcal{B}_{gX}(Q) = g \mathcal{B}_{X}(gQ), \qquad g \in
\mathbb{R}\setminus \lbrace 0 \rbrace.
\end{equation}
Let us note that this formula is identical to the one for a
complex Blue's function and complex $g$,
\begin{equation} \label{eq:kk}
B_{gX}(z) = g B_{X}(gz).
\end{equation}


\subsection{A Quaternion Blue's Function for a Non--Hermitian Random
Matrix with Hermitian and Anti--Hermitian Parts Freely
Independent} \label{s:r}

In this subsection  we apply a general formalism of the previous
part to a particular problem:  consider a non hermitian 
random $N \times N$ matrix $X$ of the form
\begin{equation} \label{eq:ll}
X = H + iH',
\end{equation}
where  the (hermitian)
random matrices $H$ and $H'$ are freely independent.
The advantage of quaternion Blue's functions 
stems from the fact, that, not-surprisingly, they
obey the   addition law
(the proof parallels the proof of an ``addition law'' for hermitized matrices
in~\cite{FZ1}, so we are not repeating it here)
\begin{equation} \label{eq:mm}
\mathcal{B}_{X}(Q) = \mathcal{B}_{H}(Q) + \mathcal{B}_{iH'}(Q) -
\frac{1}{Q}.
\end{equation}
We can now easily compute 
the quaternion Blue's  function for $X$ assuming the knowledge of
quaternion Blue's functions for $H$ and $H'$, since they 
are expressed using (\ref{eq:z}), i.e.  their complex Blue
functions.

Using  (\ref{eq:z}) and (\ref{eq:ii}), we may
rewrite (\ref{eq:mm}) in an explicit form,
\begin{displaymath}
\mathcal{B}_{X}(Q) = \beta_{H}(q, \bar{q}) 1_{2} +
\beta_{H'}(q^{\I}, \overline{q^{\I}}) \left( \begin{array}{cc} i & \\
& -i \end{array} \right)_{2} -
\end{displaymath}
\begin{equation} \label{eq:nn}
- [\beta'_{H}(q, \bar{q}) + \beta'_{H'}(q^{\I}, \overline{q^{\I}})
+ \frac{1}{\Det Q}] Q^{\dagger},
\end{equation}
where $q^{\I}$, $\overline{q^{\I}}$ are eigenvalues of $Q^{\I}$,
see (\ref{eq:c}).

This explicit form of the addition law for any ensemble
of the form $H+iH^{'}$ (where $H$ and $H^{'}$ are free) 
is one of the main results of this paper. 

Few comments are helpful.
\begin{itemize}
\item The $^{\I}$'s terms came from the multiplication formula
(\ref{eq:ii}) and the non--real number $i$ before $H^{'}$. Note that
the presence of $i$ in (\ref{eq:ll}) is essential for the
non--hermiticity and it is $i$ that brings $Q^{\I}$, so the
existence of the eigenvalues of $Q^{\I}$ on top of eigenvalues 
of $Q$ is
exactly a sign of non--hermiticity.
\item The operation $^{\I}$ changes the trace of $Q$ and  does not
change its determinant,
\begin{equation} \label{eq:oo}
\Tr Q^{\I} = i(a - \bar{a}) = -2x_{3} \neq 2x_{0} = \Tr Q, \qquad
\Det Q^{\I} = \Det Q,
\end{equation}
see (\ref{eq:c}), so that $\Tr Q^{\I}$ cannot be expressed through
$Q$'s invariants. Since in general
\begin{equation} \label{eq:pp}
q = x_{0} + i|\vec{x}| = x_{0} + i \sqrt{x^{2}_{1} + x^{2}_{2} +
x^{2}_{3}},
\end{equation}
then
\begin{equation} \label{eq:qq}
q^{\I} = -x_{3} + i \sqrt{x^{2}_{1} + x^{2}_{2} + x^{2}_{0}},
\end{equation}
and it is impossible to express $q^{\I}$ only by $q$. Note that
the eigenvalues are  related since 
\begin{equation} \label{eq:rr}
|q| = |q^{\I}|.
\end{equation}
\item The algebraic properties (\ref{eq:i}), (\ref{eq:j}) of
quaternion Green's and Blue's  functions for hermitian random matrices
remain valid for general random matrices,
\begin{equation} \label{eq:ss}
\mathcal{B}_{X}(\left( \begin{array}{cc} q & \\ & \bar{q}
\end{array} \right)_{2}) = \left( \begin{array}{cc} B_{X}(q) &
\\ & B_{X^{\dagger}}(\bar{q}) \end{array} \right)_{2} =
\left( \begin{array}{cc} B_{X}(q) & \\ & \overline{B_{X}(q)}
\end{array} \right)_{2},
\end{equation}
and identically for $\mathcal{G}_{X}$. (We use general formulae,
$B_{X^{\dagger}}(\bar{q}) = \overline{B_{X}(q)}$ and identically
for $G$.)
It is important to note, that  due to the presence of $q^{\I}$, 
\begin{equation} \label{eq:tt}
\mathcal{B}_{X}(S^{-1}QS) \neq S^{-1} \mathcal{B}_{X}(Q) S
\end{equation}
and similar relation holds  for $\mathcal{G}_{X}$. 
Indeed, this important
non--trivial behavior under the similarity transformation 
is precisely a footprint  of non--hermiticity of
$X$.
The immediate conclusion is that $(B_{X}(q)$ and
$\overline{B_{X}(q)})$ (resp. $(G_{X}(q)$ and
$\overline{G_{X}(q)})$) {\em are not} the  eigenvalues of
$\mathcal{B}_{X}(Q)$ (resp. $\mathcal{G}_{X}(Q)$).
\end{itemize}

\section{Holomorphic and Non--Holomorphic Green's Functions for a
Non--Hermitian Random Matrix with Hermitian and Anti--Hermitian
Parts Freely Independent} \label{s:u}

 In this section, we invert  functionally in a fixed point of $Q = Z$
which gives $\mathcal{G}_{X}(Z)$ and the (complex)
non--holomorphic Green's function for $X$ in particular.  
The result is given
 by a surprisingly simple and operationally convenient
formula.
We have already computed the quaternion Blue's  function for $X=H+iH^{'}$.
 We
should now invert it functionally to get the quaternion Green
function for $X$, however this is in fact unnecessary since in
this paper we are interested only in $\mathcal{G}_{X}(Z)$. Hence
we need  only to solve the equation
\begin{equation} \label{eq:xx}
\mathcal{B}_{X}(\left( \begin{array}{cc} A & i\bar{B} \\ iB &
\bar{A}
\end{array} \right)_{2}) = \left( \begin{array}{cc} z & \\ &
\bar{z} \end{array} \right)_{2},
\end{equation}
where we denote $\mathcal{G}_{X}(Z) \equiv \left(
\begin{array}{cc} A & i\bar{B} \\ iB & \bar{A} \end{array}
\right)_{2}$. We use it to extract $A = G_{X}(z, \bar{z})$ and
$-|B|^{2} = C_{X}(z, \bar{z})$.

One may ask why it is allowed to put here the regulator $\epsilon
= 0$. Note first that $\mathcal{G}_{X}(Q)$ is discontinuous in $Q
= Z$, i. e. $\mathcal{G}_{X}(Z) \neq
\mathcal{G}_{X}(Z_{\epsilon})$. On the other hand, we will show
that there are always two solutions of (\ref{eq:xx}), the first
one gives $\mathcal{G}_{X}(Z)$ and the second one,
$\mathcal{G}_{X}(Z_{\epsilon})$. Finally, the answer is that
\emph{functional inverting of a function in her discontinuity
point gives two solutions, the first one is a value of the
function in this point and the second one is a limit value of the
function when approaching the discontinuity point}. In other word,
the limit $\epsilon \to 0$ is precisely encoded in the functional
inverting.

Rewrite (\ref{eq:xx}) using (\ref{eq:nn}),
\begin{equation} \label{eq:yy}
\left( \begin{array}{cc} k + ik' - l\bar{A} & li\bar{B} \\ liB & k
- ik' - lA
\end{array} \right)_{2}) = \left( \begin{array}{cc} z & \\ &
\bar{z} \end{array} \right)_{2},
\end{equation}
where for short
\begin{eqnarray}
k & = & \beta_{H}(\mathcal{G}_{X}(Z)), \label{eq:zz}\\
k' & = & \beta_{H'}(\mathcal{G}_{X}(Z)^{\I}), \label{eq:aaa}\\
l & = & \beta'_{H}(\mathcal{G}_{X}(Z)) +
\beta'_{H'}(\mathcal{G}_{X}(Z)^{\I}) + \frac{1}{\Det
\mathcal{G}_{X}(Z)}. \label{eq:bbb}
\end{eqnarray}

Looking at the two of\mbox{}f--diagonal equations we see that this
equation has always two solutions, one with $B = 0$ and one with
$B \neq 0$.

Let us infer the first case. For $B = 0$, $\mathcal{G}_{X}(Z)$ is
diagonal, hence from (\ref{eq:ss})
\begin{equation} \label{eq:ccc}
Z = \mathcal{B}_{X}(\left( \begin{array}{cc} A & \\ & \bar{A}
\end{array} \right)_{2}) = \left( \begin{array}{cc} B_{X}(A) & \\
& \overline{B_{X}(A)} \end{array} \right)_{2},
\end{equation}
which leads to
\begin{equation} \label{eq:ddd}
A = G_{X}(z).
\end{equation}
We may call this solution the \emph{holomorphic} one since it is
expressed by a holomorphic Green's function for $X$.

The second case is a general one and we call it the
\emph{non--holomorphic} one since then $A$ is a non--holomorphic
Green's function for $X$.

To sum it up, \emph{there always exist two solutions of
(\ref{eq:xx}), the first one with $B = 0$, which reads $A =
G_{X}(z)$ and which is called the holomorphic solution and the
second one with $B \neq 0$ which is called the non--holomorphic
solution}.


\subsection{The Non--Holomorphic Solution}
\label{s:w}


Since in  (\ref{eq:zz}), (\ref{eq:aaa}), (\ref{eq:bbb}) functions $\beta$
and $\beta'$
are expressed via eigenvalues of $\mathcal{G}_{X}(Z)$ and
$\mathcal{G}_{X}(Z)^{\I}$, let us denote these eigenvalues by $g$
and $g^{\I}$, and their conjugate partners respectively. From
subsection (\ref{s:r})  we know that they are \emph{not} equal to
$G_{X}(z)$, $G_{X}(\bar{z})$.

In the non--holomorphic case the of\mbox{}f--diagonal equations of
(\ref{eq:xx}) are
\begin{equation} \label{eq:eee}
l = 0,
\end{equation}
because it is possible to divide both sides by $i\bar{B}$ or $iB$;
explicitly
\begin{equation} \label{eq:fff}
\frac{B_{H}(g) - B_{H}(\bar{g})}{g - \bar{g}} +
\frac{B_{H'}(g^{\I}) - B_{H'}(\overline{g^{\I}})}{g^{\I} -
\overline{g^{\I}}} + \frac{1}{|g|^{2}} = 0.
\end{equation}

Two diagonal equations of (\ref{eq:xx})
\begin{eqnarray*}
k + i k' - l D & = & z,\\
k - i k' - l A & = & \bar{z}
\end{eqnarray*}
gives due to (\ref{eq:eee})
\begin{eqnarray}
k & = & x, \label{eq:ggg}\\
k' & = & y, \label{eq:hhh}
\end{eqnarray}
where $z = x + iy$.  Explicitly
\begin{eqnarray}
\frac{gB_{H}(g) - \bar{g}B_{H}(\bar{g})}{g - \bar{g}} & = & x, \label{eq:iii}\\
\frac{g^{\I}B_{H'}(g^{\I}) -
\overline{g^{\I}}B_{H'}(\overline{g^{\I}})}{g^{\I} -
\overline{g^{\I}}} & = & y. \label{eq:jjj}
\end{eqnarray}

The form of these equations suggests introducing two new (complex
in general) variables, $m$ and $m'$, thus changing two equations
(\ref{eq:iii}), (\ref{eq:jjj}) into four equations
\begin{eqnarray*}
gB_{H}(g) & = & x g + m,\\
\bar{g}B_{H}(\bar{g}) & = & x\bar{g} + m,\\
g^{\I}B_{H'}(g^{\I}) & = & yg^{\I} + m',\\
\overline{g^{\I}}B_{H'}(\overline{g^{\I}}) & = &
y\overline{g^{\I}} + m'
\end{eqnarray*}
Conjugating the
first and third ones, comparing with the remaining two and
exploiting $\overline{B_{H}(z)} = B_{H}(\bar{z})$ true for
hermitian matrices, we get
\begin{equation} \label{eq:fffff}
m, m' \in \mathbb{R}
\end{equation}
and two equations
\begin{eqnarray}
B_{H}(g) & = & x + \frac{m}{g}, \label{eq:kkk}\\
B_{H'}(g^{\I}) & = & y + \frac{m'}{g^{\I}}, \label{eq:mmm}.
\end{eqnarray}
Substituting (\ref{eq:kkk}), (\ref{eq:mmm}) into the
of\mbox{}f--diagonal equation (\ref{eq:fff}) gives a 
simple result,
\begin{equation} \label{eq:ooo}
m + m' = 1.
\end{equation}

To bring the solution to the end we have to express $A$ and $B$
through $g$ and $g^{\I}$; using general expressions (\ref{eq:pp})
and (\ref{eq:qq}) we get
\begin{equation} \label{eq:rrr}
A = \Re g - i \Re g^{\I}
\end{equation}
and
\begin{equation} \label{eq:ttt}
-|B|^{2} = \frac{1}{2} [ ( (\Re g)^{2} + (\Re g^{\I})^{2} ) - (
(\Im g)^{2} + (\Im g^{\I})^{2} ) ].
\end{equation}

Finally, we can write down the main result of this section:
the algorithm of calculating a non--holomorphic Green
function for a non--hermitian random matrix $X$ with hermitian and
anti--hermitian parts freely independent.

Let us start with given  non--hermitian random matrix $X = H + iH'$,
with $H$ and $H'$ freely independent hermitian ensembles. The holomorphic
Green's or Blue's functions for $H$ and $H'$ are explicitly 
known or given implicitly by
some equations. 
The algorithm goes as follows:
\begin{enumerate}
\item  Write down two equations
\begin{eqnarray}
B_{H}(g) & = & x + \frac{m}{g}, \label{eq:uuu}\\
B_{H'}(g^{\I}) & = & y + \frac{1 - m}{g^{\I}}, \label{eq:www}
\label{eq:xxx}
\end{eqnarray}
with three unknown quantities, $g$, $g^{\I}$ and $m \in
\mathbb{R}$. Find from them $g + \bar{g}$, $g\bar{g}$, $g^{\I} +
\overline{g^{\I}}$ and $g^{\I}\overline{g^{\I}}$ expressed via
$m$.
\item Compute $m$ from the third equation,
\begin{equation} \label{eq:yyy}
g\bar{g} = g^{\I}\overline{g^{\I}}.
\end{equation}
\item Put $m$ into $g + \bar{g}$, $g^{\I} +
\overline{g^{\I}}$ and $|g|^{2}$ expressed through $m$ (step 1).
\end{enumerate} 

These three steps  yield 
 the non--holomorphic Green's function and
correlator between left and right eigenvectors for $X$:
\begin{eqnarray}
G_{X}(x, y) & = & \frac{1}{2}[ ( g + \bar{g} ) - i ( g^{\I} + \overline{g^{\I}} ) ],\label{eq:zzz}\\
C_{X}(x, y) & = & \frac{1}{4}( g^{2} + \bar{g}^{2} + (g^{\I})^{2}
+ \overline{g^{\I}}^{2} ) =\\
& = & \frac{1}{4}[ ( g + \bar{g} )^{2} + ( g^{\I} + \overline{g^{\I}} )^{2} ] - |g|^{2}.\label{eq:gggggg}
\end{eqnarray}
The general algorithm contains also a simple method of deriving an
equation (in coordinates $(x, y)$) of a borderline of $X$'s
eigenvalues' domains. The borderline is
exactly the place where holomorphic and non--holomorphic Green
functions meet together. And it is the limit $B \to 0$ that
carries the non--holomorphic solution towards its boundary with the
holomorphic one. Hence \emph{the borderline's equation} is simply
\begin{equation} \label{eq:ccccc}
B = 0,
\end{equation}
where $B$ is an appropriate element of the \emph{non--holomorphic
solution}.
This is equivalent to the  condition of vanishing of the correlator
between left and right eigenvectors,
\begin{equation} \label{eq:jjjjjj}
C_{X}(x, y) = 0,
\end{equation}
where we compute $C_{X}(x, y)$ for the \emph{non--holomorphic solution}.

Due to (\ref{eq:gggggg}) and (\ref{eq:yyy}), the 
general equation for the curve defining  the support 
of the eigenvalue domains is  given by
\begin{equation}
( g + \bar{g} )^{2} + ( g^{\I} + \overline{g^{\I}} )^{2} =
4g\bar{g},
\end{equation}
where $g + \bar{g}$, $g^{\I} + \overline{g^{\I}}$ and $g\bar{g}$
are expressed via $m$ in step 1 and explicitly in step 3, so that
this equation binds $x$ and $y$ to form the borderline.

Before closing this section, few comments are useful.
First, inferring the equations (\ref{eq:uuu}), (\ref{eq:www}) we
look for a solution that comes with its conjugate partner in pair.
(E. g., if $g$ satisfies the equation, $\bar{g}$ does it as well.) 
So that two
possibilities are allowed, either this is the pair of two \emph{
different} (and hence \emph{non--real}) numbers, or they
merge into a \emph{one real} number. The second possibility means
actually,  that $\mathcal{G}_{X}(Z)$ is
proportional (with real coefficient) to a diagonal matrix, i. e.
(a) in case of $g$, eq. (\ref{eq:uuu}), to $1_{2}$
(b) in case of $g^{\I}$, eq. (\ref{eq:www}), to $i\sigma_{3}$.
However, this is impossible since we are looking for a
non--holomorphic solution with non--diagonal $\mathcal{G}_{X}(Z)$.
To sum up, \emph{we are looking for a solution that is non--real and
comes in pair with its conjugate partner}.

Second, it is not actually our  task to find all $g$'s in this step but only
their sums, products etc. ,   e.g. we may  use the Viete
rules.

Third,  note that it is possible to use the holomorphic Green's
functions for $H$ or $H'$ rather than Blue's ones in (\ref{eq:uuu}),
(\ref{eq:www}), by inverting these equations functionally. In some
cases this approach may be considerably simpler.


\section{The Version of the Algorithm in the Case of $H$ from GUE}
\label{s:jj}

A common case is when one of the ensembles belongs to Gaussian class, 
but for a second a holomorphic Blue's function is given by a
complicated equation.
In this case one may reformulate
slightly the algorithm. 

Here let us deal with the problem of Gaussian $H$ with
\begin{equation} \label{eq:wwww}
B_{H}(s) = s + \frac{1}{s}
\end{equation}
and \emph{very complicated $B_{H'}(s)$}. If so, equations
(\ref{eq:uuu}), (\ref{eq:www}) are very complicated and another
way is preferred.

The quaternion Blue's function for $H$ is obtained easily from
(\ref{eq:z}) and reads
\begin{displaymath}
\mathcal{B}_{H}(Q) = Q + \frac{1}{Q},
\end{displaymath}
hence the equation (\ref{eq:xx}), after short manipulation, becomes
\begin{displaymath}
\mathcal{B}_{H'}(\left( \begin{array}{cc} i A & \bar{B} \\ -B & -i
\bar{A}
\end{array} \right)_{2}) = \left( \begin{array}{cc} i(A - z) &
-\bar{B} \\ B & i(\bar{z} - \bar{A}) \end{array} \right)_{2}.
\end{displaymath}
The key observation  is to invert the last equation functionally.

Let $h$, $\bar{h}$ be the eigenvalues of $\left( \begin{array}{cc}
i(A - z) & -\bar{B} \\ B & i(\bar{z} - \bar{A}) \end{array}
\right)_{2}$, i. e. from the general formula (\ref{eq:pp})
\begin{displaymath}
h = y - \Im A + i \sqrt{ (\Re A - x)^{2} + |B|^{2} } ),
\end{displaymath}
where we recall $z = x + iy$.

Hence from (\ref{eq:w})
\begin{equation} \label{eq:xxxx}
\left( \begin{array}{cc} i A & \bar{B} \\ -B & -i \bar{A}
\end{array} \right)_{2} = \gamma_{H'}(h, \bar{h})1_{2} - \gamma'_{H'}(h, \bar{h})
\left( \begin{array}{cc} i(\bar{z} - \bar{A}) & \bar{B} \\ -B &
i(A - z) \end{array} \right)_{2}
\end{equation}
which is  the equation to solve.

The of\mbox{}f--diagonal equations of (\ref{eq:xxxx}) are thus
simply
\begin{equation} \label{eq:yyyy}
\gamma'_{H'}(h, \bar{h}) = \frac{G_{H'}(h) - G_{H'}(\bar{h})}{h -
\bar{h}} = -1.
\end{equation}
The diagonal ones
\begin{displaymath}
i A = \gamma_{H'}(h, \bar{h}) - \gamma'_{H'}(h, \bar{h}) i
(\bar{z} - \bar{A}),
\end{displaymath}
become due to (\ref{eq:yyyy})
\begin{displaymath}
i A = \gamma_{H'}(h, \bar{h}) + i (\bar{z} - \bar{A}),
\end{displaymath}
or equivalently
\begin{eqnarray}
\gamma_{H'}(h, \bar{h}) = \frac{hG_{H'}(h) -
\bar{h}G_{H'}(\bar{h})}{h - \bar{h}} & = & -y,\label{eq:zzzz}\\
A + \bar{A} & = & x.\label{eq:aaaaa}
\end{eqnarray}
Note dif\mbox{}ferences between (\ref{eq:zzzz}), (\ref{eq:aaaaa})
and (\ref{eq:iii}), (\ref{eq:jjj}).
We solve the second diagonal equation, (\ref{eq:aaaaa}), by
introducing a new unknown quantity $n$,
\begin{equation} \label{eq:bbbbb}
A = \frac{x}{2} + n.
\end{equation}
The of\mbox{}f--diagonal equation (\ref{eq:yyyy}) is solved by
introducing a new unknown variable $m$,
\begin{equation} \label{eq:ddddd}
G_{H'}(h) = -h + m, \qquad m \in \mathbb{R}.
\end{equation}
Inserting this into the first diagonal equation (\ref{eq:zzzz}) we
get easily $m$,
\begin{equation} \label{eq:eeeee}
m = h + \bar{h} - y;
\end{equation}
but $h + \bar{h} = 2y - 2n$, so that
\begin{displaymath}
m = y - 2n \qquad \longrightarrow \qquad n = \frac{y - m}{2}
\end{displaymath}
and hence
\begin{equation} \label{eq:ffffff}
A = \frac{z - im}{2}.
\end{equation}
Finally, it is now possible to write down the algorithm of
calculating a non--holomorphic Green's function for a non--hermitian
random matrix $X$ with hermitian and anti--hermitian parts freely
independent and with hermitian part being a GUE random matrix.

We consider a  non--hermitian random matrix $X = H + iH'$,
with $H$ and $H'$ freely independent and $B_{H}(s) = s +
\frac{1}{s}$, is given. The complex holomorphic Green's or Blue
functions for $H'$ is known (or more often) given by some
(complicated) equation. The algorithm reads: 
\begin{enumerate}
\item Write down an equation
\begin{equation} \label{eq:hhhhh}
G_{H'}(h) = -h + m,
\end{equation}
with two unknown quantities, $h$ and $m \in \mathbb{R}$. Find from
it $h + \bar{h}$ and $h\bar{h}$ expressed via $m$.
 \item Compute $m$ from the second equation,
\begin{equation} \label{eq:jjjjj}
h + \bar{h} = y + m.
\end{equation}
\item (Necessary only for $C_{X}$ and the borderline's
equation.) Put $m$ into the expression for $h\bar{h}$.
\end{enumerate}
The results of these three steps are given explicitly by 
\begin{eqnarray}
G_{X}(x, y) & = & \frac{z - i m}{2},\label{eq:kkkkk}\\
C_{X}(x, y) & = & \frac{1}{4}[x^{2} + (y + m)^{2}] -
h\bar{h}.\label{eq:iiiii}
\end{eqnarray}
 The borderline's equation is
\begin{equation}
4h\bar{h} = x^{2} + (y + m)^{2},
\end{equation}
where $m$ is known from step 2 and $h\bar{h}$ from step 3.


\subsection{Examples}
\label{s:y}

To demonstrate the usefulness of the advocated approach, 
we recalculate three classical examples well  known in the literature.
We are not describing these models in detail, neither their 
physical context. We start with known 
holomorphic Green's (or Blue's)  functions for hermitian and
anti--hermitian parts,  referring for further details  to the
original papers.\\ 

\noindent
$\bullet$ {\bf The Girko--Ginibre Model}~\cite{GinGir,GIRKO}.\\
Gaussian complex ensemble, known also as a Girko-Ginibre ensemble, 
 corresponds to the case when $X=H+iH^{'}$, where both 
$H$ and $H^{'}$ are Gaussian Unitary Ensembles. 
Since the holomorphic Green's function in this case 
equals simply to $G(z)=(z-\sqrt{z^2-4})/2$, with imaginary part 
yielding a seminal Wigner semicircle~\cite{wigrise}, the functional inverses
of the resolvents are simply
\begin{equation} \label{eq:bbbb}
B_{H}(s) = rs + \frac{1}{s}, \qquad B_{H'}(s) = r's + \frac{1}{s},
\end{equation}
where $r$, $r'$ are real and positive constants (arbitrary variances
of the Gaussian distributions).
We apply the general algorithm:\\
Step 1: Equations (\ref{eq:uuu}), (\ref{eq:www}) are quadratic
\begin{displaymath}
rg^{2} - xg + 1 - m = 0, \qquad r'(g^{\I})^{2} - y(g^{\I}) + m =
0,
\end{displaymath}
hence from Viete rules
\begin{displaymath}
g + \bar{g} = \frac{x}{r}, \qquad |g|^{2} = \frac{1 - m}{r},
\end{displaymath}
\begin{displaymath}
g^{\I} + \overline{g^{\I}} = \frac{y}{r'}, \qquad
|g^{\I}|^{2} = \frac{m}{r'}.
\end{displaymath}
Step 2: The equation for $m$ is
\begin{displaymath}
\frac{1 - m}{r} = \frac{m}{r'} \qquad \longrightarrow \qquad m =
\frac{r'}{r + r'}.
\end{displaymath}
Step 3:  We get  finally
\begin{eqnarray}
G_{X}(x, y) & = & \frac{1}{2} ( \frac{x}{r} - i \frac{y}{r'} ),\label{eq:cccc}\\
C_{X}(x, y) & = & \frac{1}{4} ( \frac{x^{2}}{r^{2}} + \frac{y^{2}}{r'^{2}} ) - \frac{1}{r + r'}\label{eq:hhhhhh}
\end{eqnarray}
or in the special case of $r = r' = 1$,
\begin{eqnarray}
G_{X}(x, y) & = & \frac{1}{2} \bar{z},\label{eq:dddd}\\
C_{X}(x, y) & = & \frac{1}{4}( x^{2} + y^{2} ) - \frac{1}{2}.\label{eq:iiiiii}
\end{eqnarray}

Borderline  is given  by (\ref{eq:hhhhhh})
\begin{equation} 
\frac{x^{2}}{r^{2}} + \frac{y^{2}}{r'^{2}} = \frac{4}{r + r'},
\end{equation}
which  is the ellipse with semi--axes
$\frac{2r}{\sqrt{r^{2}+r'^{2}}}$ and
$\frac{2r'}{\sqrt{r^{2}+r'^{2}}}$. In the case of $r = r' = 1$
this is a circle with radius $\sqrt{2}$.
Applying Gauss law to nonholomphic solution yields spectral 
distribution (here a constant).\\

\noindent
$\bullet${\bf Model of
the Chaotic Resonance Scattering}~\cite{HAAKE}\\
Non-hermitian ensemble $X=H+iH^{'}$ is defined 
by $H$ belonging to GUE (or GOE) and $H^{'}$ is a  Wishart ensemble.
Corresponding Blue's functions were e.g. calculated
in~\cite{JNPZ1,USEVECT}.
\begin{equation} \label{eq:eeee}
B_{H}(s) = s + \frac{1}{s}, \qquad B_{H'}(s) = -\frac{cr}{1 + cs} + \frac{1}{s},
\end{equation}
where $r$, $c$ are real constants. We apply the general algorithm.\\
Step 1:
\begin{displaymath}
g^{2} - x g + 1 - m = 0, \qquad y c(g^{\I})^{2} + (c r - c m +
y)g^{\I} - m = 0,
\end{displaymath}
hence from Viete rules
\begin{displaymath}
g + \bar{g} = x, \qquad |g|^{2} = 1 - m,
\end{displaymath}
\begin{displaymath}
g^{\I} + \bar{g^{\I}} = \frac{c m - c r - y}{y c}, \qquad
|g^{\I}|^{2} = - \frac{m}{y c}.
\end{displaymath}
Step 2: The equation for $m$ is
\begin{displaymath}
1 - m = - \frac{m}{y c} \qquad \longrightarrow \qquad m = \frac{y
c}{y c - 1}.
\end{displaymath}
Step 3 gives
\begin{displaymath}
g + \bar{g} = x, \qquad g^{\I} + \bar{g^{\I}} = \frac{c}{y c -
1} - \frac{r}{y} - \frac{1}{c}, \qquad |g|^{2} = \frac{1}{1 - y c},
\end{displaymath}
thence
\begin{eqnarray}
G_{X}(x, y) & = & \frac{1}{2}[ x - i ( \frac{c}{yc - 1} -
\frac{r}{y}
- \frac{1}{c} ) ],\label{eq:gggg}\\
C_{X}(x, y) & = & \frac{1}{4}[ x^{2} + ( \frac{c}{yc - 1} -
\frac{r}{y} - \frac{1}{c} )^{2} ] - \frac{1}{1 - y
c}.\label{eq:kkkkkk}
\end{eqnarray}
The borderline's equation is
\begin{equation} \label{eq:llllll}
x^{2} + ( \frac{c}{yc - 1} - \frac{r}{y} - \frac{1}{c} )^{2} = \frac{4}{1 - y c}.
\end{equation}

\noindent
$\bullet${\bf The Complex Pastur Model}\\
By complex Pastur we mean a class $X=H+iH^{'}$, where 
$H$ is deterministic and $H^{'}$ belongs to GUE.
We choose for simplicity the deterministic part with only two
distinct opposite eigenvalues, 
so 
\begin{equation} \label{eq:iiii}
G_{H}(s) = \frac{1}{2}\left[\frac{1}{s-\mu} +\frac{1}{s+\mu}\right],
\end{equation}
where $\mu$ is a real constant. The general algorithm is being applied.\\
Step 1:
\begin{displaymath}
g = \frac{x + \frac{m}{g}}{(x + \frac{m}{g})^{2} - \mu ^{2}}
\qquad \longrightarrow \qquad (x^{2} - \mu ^{2})g^{2} + x(2m - 1)x
+ m(m - 1) = 0,
\end{displaymath}
\begin{displaymath}
(g^{\I})^{2} - yg^{\I} + m = 0,
\end{displaymath}
hence from Viete rules
\begin{displaymath}
g + \bar{g} = \frac{x(1 - 2m)}{x^{2} - \mu ^{2}}, \qquad
|g|^{2} = \frac{m(m  - 1)}{x^{2} - \mu ^{2}},
\end{displaymath}
\begin{displaymath}
g^{\I} + \bar{g^{\I}} = y, \qquad |g^{\I}|^{2} = m.
\end{displaymath}
Step 2: The equation for $m$ reads
\begin{displaymath}
\frac{m(m  - 1)}{x^{2} - \mu ^{2}} = m \qquad \longrightarrow
\qquad m = x^{2} - \mu ^{2} + 1 \qquad \mathrm{or} \qquad m = 0.
\end{displaymath}
Step 3 gives
\begin{displaymath}
g + \bar{g} = -2x - \frac{x}{x^{2} - \mu^{2}},
\qquad g^{\I} + \bar{g^{\I}} = y, \qquad |g|^{2} = x^{2} - \mu ^{2} + 1
\end{displaymath}
or
\begin{displaymath}
g + \bar{g} = \frac{x}{x^{2} - \mu ^{2}}, \qquad g^{\I} + \bar{g^{\I}} = y,
\qquad |g|^{2} = 0.
\end{displaymath}
In the second case  the first and third equation contradict each other,
so that finally
\begin{eqnarray}
G_{X}(x, y) & = & - \frac{iy}{2} - x - \frac{x}{2(x^{2} - \mu ^{2})},\label{eq:jjjj}\\
C_{X}(x, y) & = & \frac{1}{4}[ ( 2x + \frac{x}{2(x^{2} - \mu ^{2})} )^{2}
+ y^{2} ] - ( x^{2} - \mu^{2} + 1 ).\label{eq:mmmmmm}
\end{eqnarray}
The borderline's equation reads
\begin{equation} \label{eq:nnnnnn}
y^{2} = 4( x^{2} - \mu^{2} + 1 ) - ( 2x + \frac{x}{2(x^{2} - \mu ^{2})} )^{2}
\end{equation}
or after some manipulations reducing it to the form calculated first
by Stephanov~\cite{CHEM}\footnote{Original calculation
involved {\em chiral} GUE, but in the leading $N$ one-point Green's
function does not feel the difference in the bulk of the spectra. 
Note, however, that 
two-point Green's functions are different for chiral and non-chiral
complex Pastur ensembles~{\protect\cite{USPRL}}.} 
\begin{equation} \label{eq:nnnnnn1}
y^{2} = \frac{-4\mu^{2}x^{4} + x^{2}( 8\mu^{4} - 4\mu^{2} - 1 ) + 4\mu^{4}
( 1 - \mu^{2} )}{( x^{2} - \mu^{2} )^{2}}.
\end{equation}


\section{Summary}
\label{s:dd}

In this paper we  introduced the concept of {\em quaternion}
Green's function for non-hermitian ensembles with {\em complex} spectra, 
borrowing from  the analogy of {\em complex} Green's function for 
hermitian ensembles with {\em real} spectra.
The resulting  matrix resolvent encodes important spectral properties.
First, there are always two solutions for the diagonal elements (e.g. 
$\GG_{11}$). The nonholomorphic one yields, via Gauss law, 
the spectral density. The second solution is holomorphic.
Both solutions match along the spectral curve, defining the support
of the eigenvalues. The product of off-diagonal elements
gives~\cite{USEVECT} the correlator between the left and right eigenvectors, 
hence the shape  of the spectral curve can be 
also inferred by equating the value of  this correlator to zero.
The functional inverses of the quaternion Green's function, named 
quaternion Blue's function, obey for statistically independent ensembles
an addition law, following precisely similar laws in Free Random Variables 
calculus  
for the case of hermitian ensembles.   
All these  results confirm and generalize previous observations
made recently in the literature. In particular, construction 
~\cite{JNPZ1,JNPZ2} can be viewed as a special case of the quaternion. 
The similar construction proposed in~\cite{FZ1,FZ2,CW} can be viewed
as a variant of the presented here, with quaternion replaced 
by a two-by-two {\em hermitian} matrix. 
The advantage of the approach presented here lies in the fact, 
that the defined procedure is purely algebraic, and does not 
involve any additional implicit insights from e.g. diagrammatic approaches, 
used in the other works. 
It is also simpler at the operational level then the hermitization
procedure. 
Last but not least, we dare to say that 
the quaternion extension in the case of complex spectra is an esthetic 
construction.

We proposed the general, simple, algebraic algorithm of
finding the matrix--valued Green's function for an arbitrary non--hermitian
random matrix models with hermitian and anti--hermitian parts
freely independent. This  reduced the problem to solving some given
explicitly (usually polynomial) equations.
The resulting algorithm is surprisingly simple, as we hope the reader
does  infer by comparing  three classical examples recalculated here 
using our method to original derivations. 

We concentrated here on mathematical aspects and the clarity of 
the  presentation,
 postponing 
announcing new results for particular physical models  to the 
approaching papers.

We considered here only random  non-hermitian ensembles
with complex entries, 
not discussing the real or quaternion ensembles.
In the case of one-point function and in the leading large $N$ expansion, 
the Green's functions for similar  real or quaternion cases 
are expected to be similar modulo trivial rescalings. 
However, challenging problem is the subleading behavior, or 
two-point Green's functions (wide-correlators) for nonhermitian ensembles. 
We expect that an  extension of  
the quaternion technique may considerably  simplify the calculations 
of wide-correlators for two- and more-point Green's functions
for non-hermitian ensembles. 
These ideas will be addressed next.

\section*{Acknowledgments}
\label{s:ee}
This work was partially 
supported by the Polish State Committee for Scientific Research
(KBN) grant  2P03B08225 (2003-2006).
 The authors would like
to thank Jurek Jurkiewicz, Piotr  \'{S}niady and Roland Speicher  
for  valuable remarks.


\appendix


\section{Behavior of Quaternion Green's and Blue's
Functions for a Hermitian Random Matrix under Similarity
Transformations}
\label{s:kk}

In this appendix we present proofs of theorems (\ref{eq:k}) and
(\ref{eq:p}).


\subsection{Proof of (\ref{eq:k})}

Since $H$ is hermitian,
\begin{equation} \label{eq:l}
H^{\D} = \left( \begin{array}{cc} H & \\ & H \end{array}
\right)_{2N}.
\end{equation}
The diagonal structure and the equality of the diagonal elements
implies that $H^{\D}$ commutes with any matrix of the form
$S^{\U}$,
\begin{equation} \label{eq:m}
[H^{\D}, S^{\U}] = 0;
\end{equation}
it is easy to check. Hence
\begin{equation} \label{eq:n}
S^{-1, \U}H^{\D}S^{\U} = H^{\D}.
\end{equation}

Now let us choose $S$ to be a similarity transformation that is
internal in the quaternion space, i. e. if $Q$ is a quaternion,
then $S^{-1}QS$ too. There exist such transformations
as is shown in appendix \ref{s:ll}.

Now from the definition (\ref{eq:f}) and this simple property of
$H^{\D}$, (\ref{eq:n}), we have
\begin{eqnarray*}
\mathcal{G}_{H}(S^{-1}QS) & = & \frac{1}{N} \langle \bTr{2}
[(S^{-1}QS)^{\U} - H^{\D}]^{-1} \rangle = \nonumber \\ & = &
\frac{1}{N} \langle \bTr{2} [S^{-1, \U}(Q^{\U} -
H^{\D})S^{\U}]^{-1} \rangle = \ldots
\end{eqnarray*}
The $S^{\U}$s can be get out of the block--trace (and the average
of course) because of the simple relation for a block--trace,
\begin{equation} \label{eq:o}
\bTr{2} ( Y M^{\U} ) = (\bTr{2} Y) M
\end{equation}
true for any $2N \times 2N$ matrix $Y$ and $2 \times 2$ matrix
$M$. Hence
\begin{displaymath}
\ldots = S^{-1} \frac{1}{N} \langle \bTr{2} (Q^{\U} - H^{\D})^{-1}
\rangle S = S^{-1} \mathcal{G}_{H}(Q) S,
\end{displaymath}
what was to prove.


\subsection{Proof of (\ref{eq:p})}

From (\ref{eq:k}) and from the definition (\ref{eq:h}) we have
\begin{displaymath}
\mathcal{G}_{H}(S^{-1}QS) = S^{-1} \mathcal{G}_{H}(Q) S
\end{displaymath}
\begin{displaymath}
S^{-1}QS = \mathcal{B}_{H}( S^{-1} \mathcal{G}_{H}(Q) S )
\end{displaymath}
\begin{displaymath} S^{-1}\mathcal{B}_{H}(Q)S = \mathcal{B}_{H}(
S^{-1} \mathcal{G}_{H}(\mathcal{B}_{H}(Q)) S ) = \mathcal{B}_{H}(
S^{-1}QS),
\end{displaymath}
what was to prove.


\section{Quaternion Green's and Blue's Functions for a
Hermitian Random Matrix}
\label{s:ll}

In this appendix we present proofs of theorems (\ref{eq:w}) and
(\ref{eq:z}) and also some additional information.


\subsection{Proof of (\ref{eq:w}) and (\ref{eq:z})}

The proof comes in three parts, first for a given $Q$ we compute
the similarity transformation $S$ which diagonalizes $Q$, second
we invert it to get $S^{-1}$ and third we use these explicit forms
to apply (\ref{eq:i}) and (\ref{eq:k}) to get the final result.

Let us assume, that $Q = \left( \begin{array}{cc} a & i\bar{b}
\\ ib & \bar{a} \end{array} \right)_{2}$ is non--degenerate
(eigenvalues are $q \neq \bar{q}$) since we know the degenerate
case be trivial, $Q = x_{0}1_{2}$. Let us assume also that $Q$ is
not diagonal, which means $b \neq 0$, since the opposite case does
not need any diagonalizing transformation.

First let us find an explicit expression for $S$ assuming the
knowledge of $a$ and $b$. A short calculation shows that
\begin{displaymath}
S = \left( \begin{array}{cc} s_{1} & s_{1}\frac{a - q}{-ib} \\
s_{2}\frac{\bar{a} - \bar{q}}{-i\bar{b}} & s_{2} \end{array} \right)_{2}
\end{displaymath}
for any complex $s_{1}$, $s_{2}$. This is a general similarity
transformation that is internal in the quaternion space. Let us
choose for simplicity e. g. $s_{1} = ib$, $s_{2} = i\bar{b}$,
which gives
\begin{equation} \label{eq:s}
S = \left( \begin{array}{cc} ib & q - a \\ \bar{q} - \bar{a} & i\bar{b}
\end{array} \right)_{2}.
\end{equation}
(The $21$ element $\bar{q} - \bar{a}$ can be exchanged by $a - q$
due to $a + \bar{a} = \Tr Q = q + \bar{q}$.)

A determinant reads after a simple calculation
\begin{equation} \label{eq:u}
\Det S = (q - a)(q - \bar{q});
\end{equation}
it is always non--zero ($S$ is invertible) in the considered situation.

The expression for $S^{-1}$ will obviously be needed; easily
\begin{equation} \label{eq:v}
S^{-1} = \frac{1}{q - \bar{q}} \left( \begin{array}{cc}
\frac{i\bar{b}}{q - a} & -1 \\ 1 & \frac{ib}{q - a} \end{array}
\right)_{2}.
\end{equation}

Finally
\begin{displaymath}
\mathcal{G}_{H}(Q) = \mathcal{G}_{H}( S^{-1} \left(
\begin{array}{cc} q & \\ & \bar{q} \end{array} \right)_{2} S ) =
\end{displaymath}
\begin{displaymath}
= \frac{1}{q - \bar{q}} \left( \begin{array}{cc}
\frac{i\bar{b}}{q - a} & -1 \\ 1 & \frac{ib}{q - a} \end{array}
\right)_{2} \left( \begin{array}{cc} G_{H}(q) & \\ & G_{H}(\bar{q})
\end{array} \right)_{2} \cdot
\end{displaymath}
\begin{displaymath}
\cdot \left( \begin{array}{cc} ib & q - a \\ \bar{q} - \bar{a} & i\bar{b}
\end{array} \right)_{2} = \ldots
\end{displaymath}
Noticing also $-|b|^{2} = (a - q)(\bar{a} - q)$ and multiplying
these three matrices we immediately recover the final result
(\ref{eq:w}),
\begin{displaymath}
\ldots = \frac{qG_{H}(q) - \bar{q}G_{H}(\bar{q})}{q - \bar{q}}
1_{2} - \frac{G_{H}(q) - G_{H}(\bar{q})}{q - \bar{q}} Q^{\dagger}.
\end{displaymath}

The identical construction leads to (\ref{eq:z}).


\subsection{Some Expressions Helpful to Cross--Check
(\ref{eq:w}) and (\ref{eq:z})}
\label{s:l}

Let us note the following easy to check relations,
\begin{eqnarray}
\gamma_{H}(B_{H}(q), B_{H}(\bar{q})) & = & \frac{\beta_{H}(q, \bar{q})}
{\beta'_{H}(q, \bar{q})},  \label{eq:cc}\\
\gamma'_{H}(B_{H}(q), B_{H}(\bar{q})) & = &
\frac{1}{\beta'_{H}(q, \bar{q})} \label{eq:dd}
\end{eqnarray}
and similarly in the opposite way.

Moreover, all $\gamma$s and $\beta$s are real.

Using these expressions we can e. g. immediately show the
cross--check relation $\mathcal{G}_{H}(\mathcal{B}_{H}(Q)) =
\mathcal{B}_{H}(\mathcal{G}_{H}(Q)) = Q$ which ensures us that the
derivation is correct.


\section{Quaternion Green's and Blue's Functions for a
General Random Matrix Multiplied by a Fixed Complex Number}
\label{s:mm}

From the definition (\ref{eq:f}) and for complex $g \neq 0$:
\begin{displaymath}
\mathcal{G}_{gX}(Q) = \frac{1}{N} \langle \bTr{2} (Q^{\U} - (g
X)^{\D})^{-1} \rangle =
\end{displaymath}
\begin{displaymath}
= \frac{1}{N} \langle \bTr{2} (Q^{\U} - \left( \begin{array}{cc} g
X & \\ & \bar{g}X^{\dagger}
\end{array} \right)_{2N})^{-1} \rangle =
\end{displaymath}
\begin{displaymath}
= \frac{1}{N} \langle \bTr{2} (Q^{\U} - \left( \begin{array}{cc}
g1_{N} & \\ & \bar{g}1_{N} \end{array} \right)_{2N} X^{\D})^{-1}
\rangle =
\end{displaymath}
\begin{displaymath}
=\frac{1}{N} \langle \bTr{2} ((\left( \begin{array}{cc}
\frac{1}{g}1_{N} & \\ & \frac{1}{\bar{g}}1_{N} \end{array}
\right)_{2N} Q^{\U} - X^{\D})^{-1} \left( \begin{array}{cc}
\frac{1}{g}1_{N} & \\ & \frac{1}{\bar{g}}1_{N} \end{array}
\right)_{2N}) \rangle =
\end{displaymath}
\begin{displaymath}
= \mathcal{G}_{X}( \left(
\begin{array}{cc} 1/g & \\ & 1/\bar{g} \end{array}
\right)_{2} Q ) \left( \begin{array}{cc} 1/g & \\ & 1/\bar{g}
\end{array} \right)_{2} = \ldots
\end{displaymath}
this is the formula we were looking for,
\begin{equation} \label{eq:1}
\ldots = \mathcal{G}_{X}( \left( \begin{array}{cc} 1/g & \\
& 1/\bar{g} \end{array} \right)_{2} Q ) \left(
\begin{array}{cc} 1/g & \\ & 1/\bar{g}
\end{array} \right)_{2}.
\end{equation}






\begin{thebibliography}{99}
\bibitem{JNPZ1} R. A. Janik, M. A. Nowak, G. Papp, J. Wambach,
and I. Zahed, Phys. Rev. E \textbf{55} (1997) 4100.

\bibitem{JNPZ2}
 R. A. Janik, M. A.
Nowak, G. Papp and  I. Zahed, Nucl. Phys. \textbf{B501} (1997) 603.

\bibitem{FZ1}
J. Feinberg and A. Zee,
        {\it Nucl. Phys.} {\bf B501} (1997) 643.

\bibitem{FZ2}
J. Feinberg and A. Zee,
        {\it Nucl. Phys.} {\bf B504} (1997) 579.

\bibitem{CW}
J.T. Chalker and Z. Jane Wang,
        {\it Phys. Rev. Lett.} {\bf 79} (1997) 1797.

\bibitem{FS}
Y.V. Fyodorov and H.-J.Sommers,
        {\it unpublished}.

\bibitem{VOICULESCU}
D. Voiculescu, 
        Invent. Math. {\bf 104} (1991) 201;
D.V. Voiculescu, K.J. Dykema and A. Nica, 
        {\em Free Random Variables},
        (Am. Math. Soc., Providence, RI, 1992).

\bibitem{SPEICHER}
R. Speicher, Math. Ann. {\bf 298} (1994) 611.

\bibitem{ZEE} A. Zee, Nucl. Phys. \textbf{B474} (1996) 726.

\bibitem{GUHR}
T. Guhr, A. Mueller-Groeling and H.A. Weidenmueller, Phys. Rep. {\bf
  299}
(1998) 189 and references therein. 

\bibitem{GENERALHER}
see e.g. M.L. Mehta, {\it Random matrices} (Academic Press, New York,
        1991);
C.E. Porter, {\it Statistical Theories of Spectra: Fluctuations}
        (Academic Press, New York, 1969).

\bibitem{ZINNJUSTIN}
see P.~Di~Francesco, P.~Ginsparg and J.~Zinn-Justin, 
        Phys.~Rept. {\bf 254} (1995)~1, and references therein.

\bibitem{HAAKE}
F. Haake et al.,
        {\it Zeit. Phys.} {\bf B88} (1992) 359;\\
N. Lehmann, D. Saher, V.V. Sokolov and H.-J. Sommers,
        {\it Nucl. Phys.} {\bf A582} (1995) 223.

\bibitem{CHEM} M. A. Stephanov, Phys. Rev. Lett. \textbf{76} (1996) 4472;
G. Akemann, J. Phys. {\bf A36} (2003) 3363.

\bibitem{THETA}
R.A. Janik, M.A. Nowak, G. Papp and I. Zahed, Avta Phys. Pol. {\bf B32}(2001)
1297.

\bibitem{HATANO}
N. Hatano and D.R. Nelson, Phys. Rev. Lett. {\bf 77} (1966) 570;
J. Feinberg and A. Zee, Phys. Rev. {\bf E59} (1999) 6433;
I. Ya. Goldsheid and B.A. Khoruzhenko, Phys. Rev. Lett. {\bf 80} (1998) 2897;
R.A. Janik, M.A. Nowak, G. Papp and I. Zahed, Acta Phys. Pol. {\bf B30} (1999) 45.

\bibitem{EWA}
E. Gudowska-Nowak, G. Papp and J. Brickmann, Chem. Phys. {\bf 232}
(1998) 247.

\bibitem{OURRECENT}
E. Gudowska-Nowak, R.A. Janik, J. Jurkiewicz and M.A. Nowak, 
 Nucl. Phys. {\bf B 670 } (2003) 479. 

\bibitem{TEODOR}
R. Teodorescu, E. Bettelheim, O. Agam, A Zabrodin and 
P. Wiegmann, hep-ph/0401165. 





\bibitem{BZ}
E. Br\'{e}zin and A. Zee, 
        Phys. Rev. {\bf E49} (1994) 2588;
E. Br\'{e}zin and A. Zee, 
        Nucl. Phys. {\bf B453} (1995) 531.

\bibitem{GinGir} J. Ginibre,
        {\it J. Math. Phys.} {\bf 6} (1965) 440.

\bibitem{GIRKO}
V.L. Girko,
        {\it Spectral theory of random matrices} (in Russian), Nauka,
        Moscow (1988) and references therein.

\bibitem{THOOFT}
G. 't Hooft, Nucl. Phys. {\bf B75} (1974) 464



\bibitem{Bro86}
L.G. Brown, {\em Geometric methods in operator algebras}, pages 1-35, 
 Longman Sci. Tech., Harlow, (1986). 
\bibitem{HL00}
U. Haagerup and F. Larsen, J. Funct. Anal. {\bf 176(2)} (2000) 331. 
\bibitem{Sni02}
P. \'{S}niady, J. Funct. Anal. {\bf 193(2)} (2002) 291.



\bibitem{SOMMERS}
Y.V. Fyodorov and H.-J.Sommers,
        {\it J. Math. Phys.} {\bf 38} (1997) 1918;\\
Y.V. Fyodorov, B.A. Khoruzhenko  and H.-J.Sommers,
       {\it Phys. Lett.} {\bf A226} (1997) 46;\\
H.-J. Sommers, A. Crisanti, H. Sompolinsky and Y. Stein, {\it
Phys. Rev. Lett.} {\bf 60} (1988) 1895.






\bibitem{NEWPAPER}
E. Gudowska-Nowak, A. Jarosz, R. Janik, J. Jurkiewicz and M.A. Nowak,
in preparation.  
\bibitem{USEVECT}
R.A. Janik, W. Noerenberg, M.A. Nowak, G. Papp and I. Zahed,
Phys. Rev. {\bf E60} (1999) 2699.
\bibitem{MEHLIG}
J.T. Chalker abd B. Mehlig, Phys. Rev. Lett. {\bf 81} (1998) 3367.
\bibitem{SPEICHERNEU}
P. Neu and R. Speicher, J. Stat. Phys. {\bf 80} (1995) 1279. 
\bibitem{GROSS}
R.  Gopakumar and D.J. Gross, Nucl. Phys. {\bf B451}
(1995) 379.
\bibitem{BLUESHADES} R. A. Janik, M. A. Nowak, G. Papp, I. Zahed, 
Acta Phys. Polon. {\bf B28} (1997) 2947.



\bibitem{wigrise} E. Wigner, {\it Can. Math. Congr. Proc.} p.174
(University of Toronto Press) and
other papers reprinted in  C.E. Porter {\it Statistical Theories of
Spectra: Fluctuations} (Academic Press, New York, 1965);\\
M.L. Mehta, {\it Random Matrices} (Academic Press, New York, 1991).

\bibitem{USPRL}
R.A. Janik, M.A. Nowak, G. Papp and I. Zahed, Phys. Rev. Lett. {\bf 77}
(1996) 4876.  

\end{thebibliography}
\end{document}